\documentclass[aps,prc,reprint,superscriptaddress,nofootinbib,floatfix,showkeys]{revtex4-2}
\usepackage[utf8]{inputenc}
\usepackage[T1]{fontenc}

\usepackage{microtype}
\usepackage{amsmath}
\usepackage{amssymb}
\usepackage{graphicx}
\usepackage{multirow}
\usepackage{dcolumn}
\usepackage{xcolor}
\usepackage{bm}
\usepackage{comment}
\usepackage{tabularx}
\usepackage{booktabs}
\usepackage{subcaption}
\usepackage[justification=raggedright]{caption}

\usepackage[italicdiff]{physics}
\usepackage[colorlinks=True,allcolors=blue]{hyperref}
\usepackage{orcidlink}

\usepackage{soul}

\newcommand\IITRxPH{Department of Physics,
Indian Institute of Technology Roorkee, Roorkee 247667, India} 

\newcommand{\UGxICTQT}{International Centre for Quantum Technologies,
University of Gda\'nsk,
80-309 Gda\'nsk, Poland}

\newcommand\IITRxCPQCT{Centre for Photonics and Quantum Communication Technology, 
Indian Institute of Technology Roorkee, Roorkee 247667, India}

\newcommand{\UGxIFTiA}{Institute of Theoretical Physics and Astrophysics,
Faculty of Mathematics, Physics and Informatics,
University of Gda{\'n}sk, 80-308 Gda{\'n}sk, Poland}

\begin{document}

\title{$np$ spin correlations in the deuteron ground state}

\author{Ashutosh Singh\,\orcidlink{0009-0001-9202-571X}}
\email{ashutosh_s@ph.iitr.ac.in}
\affiliation{\IITRxPH}

\author{Ankit Kumar Das}
\affiliation{\IITRxPH}

\author{Ankit Kumar\,\orcidlink{0000-0003-3639-6468}}
\affiliation{\UGxICTQT}
\affiliation{\UGxIFTiA}

\author{P. Arumugam\,\orcidlink{0000-0001-9624-8024}}
\affiliation{\IITRxPH}
\affiliation{\IITRxCPQCT}


\begin{abstract}
The deuteron is the simplest atomic nucleus made of two particles — a proton and a neutron. In this work, we study how their spins are quantum entangled with each other.
We study two cases: when the deuteron is in a fixed projection of total angular momentum, and when it exists in a superposition of all projections. 
Our findings show that the spins are most entangled when the total projection is zero, and that strong entanglement still exists even when all spin states are superposed.
\end{abstract}

\maketitle

\section{Introduction}
Correlations between subsystems which do not have a classical counterpart gives us an indication of the nonclassicality of quantum states. 
The study of nonclassical correlations, particularly entanglement, has become a key topic in many areas of research~\cite{review_amico_vedralRevModPhys.80.517}. The is becuse of a crucial role entanglement plays a resource in quantum cryptography~\cite{cryptography_PhysRevLett.67.661} and quantum computing~\cite{quantum_computation_Jozsa2003-JOZOTR, quantum_computation_PhysRevLett.91.147902, quantum_computing_iontrap10.1119/1.4948608}, quantum chemistry~\cite{chemistry_https://doi.org/10.1002/qua.24898}, atomic physics~\cite{atomic_Dehesa_2012, atomic_Qvarfort_2020}, and condensed matter physics~\cite{review_many_bodyRevModPhys.82.277, condenced_LAFLORENCIE20161}. 
Thus, understanding and quantification of entanglement is a fundamental problem in quantum information theory, which has been extensively explored in the literature~\cite{review_horodecki_RevModPhys.81.865}.
For pure bipartite states, entanglement can be easily determined using Schmidt decomposition, but it becomes a complicated task for the mixed state. For mixed states, we have different separability criteria such as Peres-Horodecki criteria, entanglement negativity, and concurrence~\cite{seperability_HORODECKI1997333, qi-neg-ppt:PhysRevLett.77.1413, qi-entropy:PhysRevLett.78.2275, qi-concurr-orig:PhysRevLett.78.5022}. This quantification of entanglement is based on the choice of subsystem considered for our study. 

In the case of distinguishable particles, the notion of entanglement is based on the tensor product of the subsystem Hilbert spaces, as thoroughly discussed in~\cite{review_horodecki_RevModPhys.81.865}. However, for indistinguishable particles, defining subsystems is challenging, as the Hilbert space is no longer expressed as a tensor product of subsystem Hilbert spaces. In fermionic systems, both particle-based approaches~\cite{indistin_schelchman_PhysRevA.64.022303, indistinguishable_ECKERT200288} and mode entanglement-based approaches~\cite{indis_mode_debarba_PhysRevA.95.022325, indis_mode_gigena_PhysRevA.94.042315, indis_mode_gigena_PhysRevA.95.062320, indis_mode_shapourian_PhysRevA.99.022310} are used. Recently, more emphasis has been placed on mode entanglement, as the definition of subsystems is well-defined in this case~\cite{indis_mode_benatti_PhysRevA.89.032326, indis_mode_friis_PhysRevA.87.022338}.
Nevertheless, particle entanglement remains essential in systems where the constituents can be effectively distinguished through additional degrees of freedom, such as spin, isospin, or spatial separation. This perspective is particularly relevant in nuclear physics, where nucleons are distinguishable by their spin and isospin projections, allowing the construction of reduced density matrices in well-defined bases.

Recently, there has been a growing interest in understanding the structure and role of entanglement in many-body problems, particularly in nuclear physics~\cite{nuclear_dmrg_PhysRevC.92.051303, nuclear_lipkin_PhysRevA.100.062104, nuclear_lipkin_PhysRevA.104.032428, nuclear_Kruppa_2021, nuclear_robin_PhysRevC.103.034325, nuclear_krupa_2022_PhysRevC.106.024303, nuclear_denis_PhysRevD.106.123006, nuclear_entropy_PhysRevC.107.054308, nuclear_entropy_PhysRevC.108.054309, nuclear_fermionic_PhysRevC.107.044318, nuclear_gorton_Johnson_2023,   qi-concurr:dong2023}.
This area of nuclear entanglement  bridges quantum information science and nuclear physics, providing new tools and perspectives for both classical and quantum nuclear problem-solving.
Recent investigations into quantum correlations in the Lipkin model~\cite{nuclear_lipkin_PhysRevA.100.062104, nuclear_lipkin_PhysRevA.104.032428}, seniority model~\cite{nuclear_krupa_2022_PhysRevC.106.024303}, and two-nucleon systems~\cite{nuclear_Kruppa_2021} have made significant contributions.
Apart from a theoretical academic interest, it also has practical implications in areas of traditional shell model calculation where entanglement entropy serves as a metric for the truncation of computational model space~\cite{nuclear_gorton_Johnson_2023}.
This means fewer computational resources are needed, without losing much accuracy, making complex nuclear calculations much more efficient.

The present quantifies the spin correlations in the simplest nucleus, deuteron, which comprises a proton and a neutron that are loosely bound with a small binding energy.
Moreover, the deuteron happens to be the only two-body nucleus existing in nature.
We derive the reduced density matrix encompassing their spin degrees of freedom, which reveals a significant amount of quantum correlations in the ground state. 
These entanglement quantifiers can be further measured experimently as proposed by Dong Bai~\cite{qi-concurr:dong2023}, and hence this work could find applications in the inference of various nuclear properties.

\section{Deuteron Ground State}      \label{sec:quantification_formalism_v3}

Deuteron is the only two-nucleon bound state existing in nature. It comprises of a neutron and a proton weakly bound together in an admixture (superposition) of $\mathcal{S}$ and $\mathcal{D}$ orbital angular momentum states~\cite{deut-zhaba2017, deut-Blatt:1952ije, deut-general:LEVCHUK2000449, deut-general:PhysRevC.65.014002, deut-general:PhysRevC.82.034004, deut-general:PhysRevC.84.034003}.
Accordingly, the orbital angular momentum quantum number takes values $L = 0$ and $2$, with an experimentally known probability of the $\mathcal{S}$ state $\approx 94\%$. 
The parity of exchange interactions fixes the  spin and isospin quantum numbers as $S = 1$ and $T = 0$, with the resulant total angular momentum $J = 1$.
The wavefunction for a fixed projection $M = 0,\pm 1$ can therefore be written as:
\begin{equation}
\ket{\Psi^{(M)}} = \sum_{L}\alpha_{L} \ket{R_{L}}\ket{LS;JM}\ket{TM_{T}},
\end{equation}
where $\abs{\alpha_{L}}^2$ are the probabilities of the deuteron in different radial vectors $\ket{R_L}$.
In the decoupled basis we can utilize the Clebsh-Gordan (CG) series to substitute:
\begin{equation}
\ket{LS;JM} = \sum_{M_{L}}\sum_{M_{S}} \mqty( L& S& J  \\  M_{L}& M_{S}& M) \ket{LM_{L}}\ket{SM_{S}},
\end{equation}
where $M_L$ and $M_S$ are the $z$ projections of $L$ and $S$, respectively, and the bracketed term represents is the CG coefficient. 
In the position representation, $\ket{R_L}$ and $\ket{LM_{L}}$ are the radial wave function and the spherical harmonics, respectively:
\begin{equation}
\braket{\vec r}{R_L} = R_L(r), 
\hspace{1cm}
\braket{\vec{r}}{LM_{L}} = Y_{L}^{M_{L}}\qty(\theta, \phi),
\end{equation}
where $\vec{r} = (r,\theta,\phi)$ is the relative position vector between the two nucleons.   
We can therefore write
\begin{multline}
\ket{\Psi^{(M)}} = \sum_{L}\alpha_{L} \ket{R_{L}}\sum_{M_{L},M_{S}} \mqty( L& S& J  \\  M_{L}& M_{S}& M) \\  \ket{L,M_L} \ket{S,M_{S}}\ket{T,M_{T}}. 
\label{eq:psi_m}
\end{multline}
It is far more convenient to move to basis that labels the two nucleons as $AB$, where the subsystems $A$ and $B$ represent the neutron and the proton, respectively.

In this work, we are interested in the entanglement between neutron-proton spins, and the corresponding reduced spin density matrix is obtained by tracing out other degrees of freedom:
\begin{align}
\Hat{\rho}^{(M)}_\text{spin} =& \Tr_{R,L,T}\qty(\ketbra{\Psi^{(M)}})  \notag \\
=& \sum_{L, M_{L}}\sum_{M_{S},M_{S}'} \abs{\alpha_{L}}^2 \mqty( L& S& J  \\  M_{L}& M_{S}'& M) \mqty( L& S& J  \\  M_{L}& M_{S}& M) \notag \\
& \hspace{2cm} \ket{S,M_{S}'} \bra{S,M_{S}}.
\end{align}
Using the completeness relation, and the fact $M_{L}+ M_{S}' = M_{L}+ M_{S}$, it reduces to [see Appendix~\ref{appen:density_matrix_single_proj} for explicit derivation]
\begin{align}
\Hat{\rho}^{(M)}_\text{spin} = & \sum_{L,M_{L},M_{S}} \abs{\alpha_{L}}^2 \mqty( L& S& J  \\  M_{L}& M_{S}& M)^2 \sum_{m_{1},m_{2}} \mqty(s_{1} & s_{2} & S \\ m_{1} & m_{2} & M_{S}) \notag \\
& \sum_{m_{1}',m_{2}'} \mqty(s_{1} & s_{2} & S \\ m_{1}' & m_{2}' & M_{S}) \ket{m_{1},m_{2}} \bra{m_{1}',m_{2}'},
\label{eq:rho_final_M}
\end{align}
where, $s_1 = 1/2$ and $s_2 = 1/2$ are the nucleon spins, and $m_1$ and $m_2$ are their corresponding projections along the $z$ axis.

The analysis done upto now is useful in the case when the deuteron sits in a weak external magnetic field gradient that segregates the nuclei on the basis of $M$ values.
However, in the absence of an external magnetic field the wave function as a linear superposition of $M = 0, \pm 1$: 
The principle of equal prior probabilities implies that all projections are equally probable, i.e.,
\begin{equation}
\ket{\Psi}=\frac{1}{\sqrt{3}}\sum_{M}{e^{i\zeta_M}}\ket{\Psi^{(M)}}, \label{eq:psi_all}
\end{equation}
where $\zeta_{M} \in (0,2\pi)$ are arbitrary phases.
Since global phases do not matter, we set $\zeta_{0} = 0$ as a reference. 
Following the same procedure as before, the corresponding spin density matrix is given by
\begin{align}
\Hat{\rho}_\text{spin}
=& \frac{1}{3}\sum_{M,M'}{e^{i(\zeta_M-\zeta_{M'})}}\sum_{L}{\abs{\alpha_{L}}^2} 
\notag \\
& \sum_{M_{L},M_S,M_S'} \mqty( L & S & J \\ M_{L} & M_{S} & M )  \mqty( L & S & J \\ M_{L} & M_{S}' & M') 
\notag \\
& \sum_{m_{1},m_{2}, m_{1}',m_{2}'} \mqty(s_{1} & s_{2} & S \\ m_{1} & m_{2} & M_{S}) \mqty(s_{1} & s_{2} & S \\ m_{1}' & m_{2}' & M_{S}') 
\notag \\
& \hspace{2cm} \ket{m_{1},m_{2}} \bra{m_{1}',m_{2}'}. \label{eq:rho_all_m}
\end{align}
Given $s_1 = s_2 = 1/2$, the bipartite spin density matrix is a $4 \times 4$ hermitian operator, whose entanglement structure could be easily probed  with standard measures, e.g., entanglement entropy, mutual information, and negativity [see Appendix~\ref{appendix:entanglement_measure_formalism} for details].

\section{Spin Correlations}
\label{sec:result_and_diss}


We now show the quantification of entanglement based on the reduced density matrices derived in the preceeding section.
Following the same order, we shall first look at the case of a weak external magnetic field when where we could choose the deuteron being in a specific projection of the total angular momentum. For scientific curiousity we will later we shall later study the case of no magnetic field, where the deuteron assumes a generic superposition of all momentum projections.

\begin{figure*}[t!]
\centering

\begin{subfigure}[t]{0.48\textwidth}
\centering
\includegraphics[width=\linewidth, , trim={0 0 0 10cm}, clip]{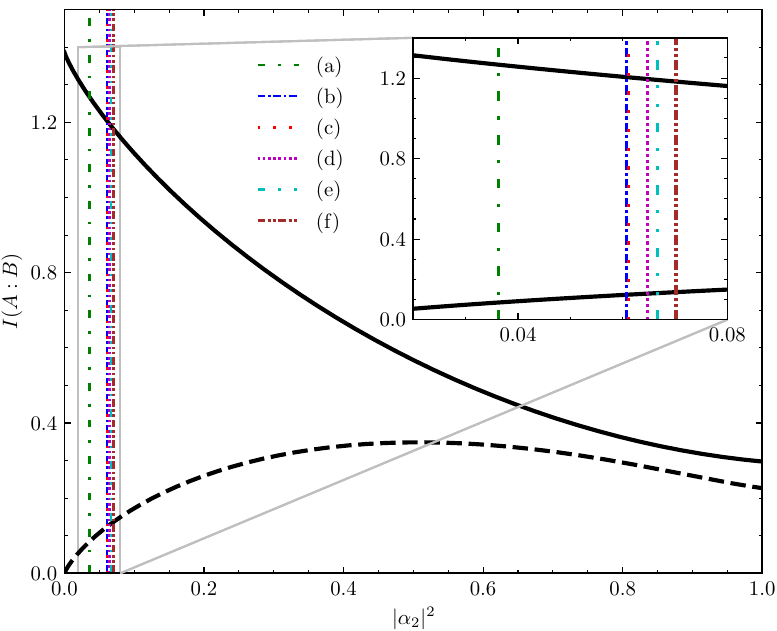}
\end{subfigure}
\hfill
\begin{subfigure}[t]{0.48\textwidth}
\centering
\includegraphics[width=\linewidth]{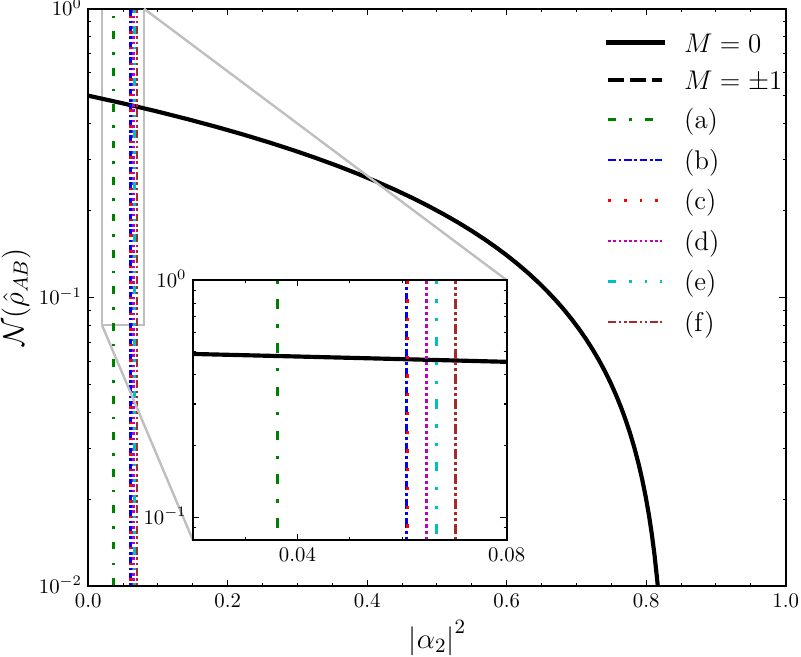}
\end{subfigure}

\caption{Bipartite mutual information $I(A\!:\!B)$ and entanglement negativity $\mathcal{N}$ of the spin correlation matrix in the deuteron ground state. The parameter $\abs{\alpha_{2}}^2$ represents the probability of $L=2$ orbital state. Vertical lines correspond to values for six different potential models listed in Table~\ref{tab:table1}. Inset: Zoomed-in region of experimentally implied range of the probability.}

\label{fig:combined_ent_mi_neg}
\end{figure*}

The deuteron ground state is mostly $L=0$, with a small admixture of $L=2$ state $(\approx 6\%)$, resulting in a mixed reduced spin density matrix.
Accordingly, in Fig.~\ref{fig:combined_ent_mi_neg} and Table~\ref{tab:table1}  we show the mutual information and entanglement negativity of the bipartite state in six different potential models that estimate the ground state.
These include 
next-to-leading order (NLO)~\cite{deut-pot-:nnlo_EPELBAUM2000295}, 
Argonne v14~\cite{deut-pot:argonne1974, deut-pot-argonne:1984PhysRevC.29.1207}, 
next-to-next-to-leading order (NNLO)~\cite{deut-pot-:nnlo_EPELBAUM2000295}, 
soft core Reid68~\cite{deut-poten:ried1968}, 
standard Woods-Saxon potential~\cite{deut-pot-:woods2014}, 
and Hamada-Johnston potential~\cite{deut-pot:HAMADA1962382}.
The spin correlations are highest in the $M=0$ case, and lowest in the $M = \pm 1$ state.
The $M=0$ state shows a monotonic decrease of mutual information with an increase in the admixture probability, while the $M=\pm 1$ state has a maxima near $\abs{\alpha_2}^2 \approx 0.5$.

\begin{table*}[hbt!]
\caption{Mutual information and entanglement negativity of the spin density matrix of deuteron in a weak external magnetic field. $|\alpha_2|^2$ denotes the probability of the $L=2$ state in different potential models. Values of parameter $\abs{\alpha_2}^2$ is obtained by reference~\cite{deut-pot-:nnlo_EPELBAUM2000295, deut-pot:argonne1974, deut-pot-argonne:1984PhysRevC.29.1207, deut-pot-:nnlo_EPELBAUM2000295, deut-poten:ried1968, deut-pot-:woods2014, deut-pot:HAMADA1962382} given in original paper of potential models applied for the calculation of deuteron wave function. These calculation are for fixed value of projection quantum number $M$.} 
\label{tab:table1}
\begin{ruledtabular}
\begin{tabular}{|c|l|c|c|c|c|c|} 
\multirow{2}{*}{Sr.} & \multirow{2}{*}{Potential} & \multirow{2}{*}{$\abs{\alpha_2}^2$ (in $\%$)}   & \multicolumn{2}{c|}{$I(A:B)$}  & \multicolumn{2}{c|}{$\mathcal{N}{\qty(\Hat{\rho}_{\text{spin}})}$} \\  \cline{4-7}
& & &  $M=0$ & $M = \pm 1$ & $M=0$ & $M = \pm 1$  \\ \hline\hline  \\[-0.5em]
a &  NLO~\cite{deut-pot-:nnlo_EPELBAUM2000295}   &  3.62   & 1.2671 & 0.0846 & 0.4784 & 0.0 \\ 
b &  Argonne v14~\cite{deut-pot:argonne1974, deut-pot-argonne:1984PhysRevC.29.1207}   &  6.08    & 1.2063 & 0.1225 & 0.4640 & 0.0  \\  
c &  NNLO~\cite{deut-pot-:nnlo_EPELBAUM2000295}  &  6.11   & 1.2039 & 0.1239 & 0.4634 & 0.0  \\ 
d &  Soft core Reid68~\cite{deut-poten:ried1968}   &  6.47    & 1.1968 & 0.1281 & 0.4616 & 0.0  \\ 
e &  Standard Wood-Saxon potential~\cite{deut-pot-:woods2014}  &  6.66  & 1.1921 & 0.1309 & 0.4604 & 0.0  \\
f & Hamada-Johnston~\cite{deut-pot:HAMADA1962382}   &   7.02    & 1.1829 & 0.1363 & 0.4580 & 0.0  \\ 
\end{tabular}
\end{ruledtabular}
\end{table*}



Negativity stands as one of the most important correlation quantifiers in information science, and hence we anticipate it to offer valuable insights into the nulear structure and reactions.
In Table~\ref{tab:table1} we show the negativity for the reduced spin density matrix . 
In a manner similar to mutual information, we have plotted the negativity as a function of parameter $\abs{\alpha_2}^2$ in the right panel of Fig.~\ref{fig:combined_ent_mi_neg}, which shows that the entanglement is significant in the $M=0$ projection, and vanishes rapidly for large $|\alpha_2|^2$.
The $M=\pm 1$ state doesn't show any significant amount of negativity at all, which clearly agress with the results of the mutual information. 
For the deuteron we have $|\alpha_2|^2 \approx 0.06$ (vertical lines in Fig.~\ref{fig:combined_ent_mi_neg}), and hence the the two nucleons are close to maximally entangled in their spins.


\begin{table*}[hbt!!]
\caption{Similar to table~\ref{tab:table1} the entanglement quantifiers for different potential models is calculated for the case considering all possible projection of $J$ as equally probable. $\zeta_{\pm 1}=0,\pi$ representing the sign of coefficient in Eq.~\eqref{eq:rho_all_m} to $\pm1$.}
\label{tab:table2}
\begin{ruledtabular}
\begin{tabular}{|c|l|c|c|c|c|c|} 
\multirow{2}{*}{Sr.} & \multirow{2}{*}{Potential} & \multirow{2}{*}{$\abs{\alpha_2}^2$ (in $\%$)}   & \multicolumn{2}{c|}{$I(A\!:\!B)$}  & \multicolumn{2}{c|}{$\mathcal{N}{\qty(\Hat{\rho}_{\text{spin}})}$}  \\  \cline{4-7}
& & &  $\zeta_{+1} \ne \zeta_{-1}$ & $\zeta_{+1} = \zeta_{-1}$ & $\zeta_{+1} \ne \zeta_{-1}$ & $\zeta_{+1} = \zeta_{-1}$   \\ \hline
a &  NLO~\cite{deut-pot-:nnlo_EPELBAUM2000295}   &  3.62   & 1.2671 & 0.2615 & 0.4784 & 0.1558  \\ 
b &  Argonne v14~\cite{deut-pot:argonne1974, deut-pot-argonne:1984PhysRevC.29.1207}   &  6.08    & 1.2063 & 0.2747 & 0.4640 & 0.1486  \\ 
c &  NNLO~\cite{deut-pot-:nnlo_EPELBAUM2000295}  &  6.11   & 1.2039 & 0.2753 & 0.4634 & 0.1483  \\ 
d &  Soft core Reid68~\cite{deut-poten:ried1968}   &  6.47    & 1.1968 & 0.2770 & 0.4616 & 0.1474  \\ 
e & Standard Wood-Saxon potential~\cite{deut-pot-:woods2014}  &  6.66  & 1.1921 & 0.2781 & 0.4604 & 0.1468  \\ 
f & Hamada-Johnston~\cite{deut-pot:HAMADA1962382}   &   7.02    & 1.1829 & 0.2803 & 0.4580 & 0.1456  \\ 
\end{tabular}
\end{ruledtabular}
\end{table*}

In the absence of a weak external magnetic field all $M$ projections are degenerate. Accorindgly, the quantum state is an equal coherent superposition of all $M$ projections, with the reduced spin density matrix given by Eq.~\eqref{eq:rho_all_m}.
To incorporate all possible (relative) phases corresponding to different projections, we introduced two additional parameters, $\zeta_{\pm 1}$.


\begin{figure*}[t!]
\centering

\begin{subfigure}[t]{0.48\textwidth}
\centering
\includegraphics[width=\linewidth]{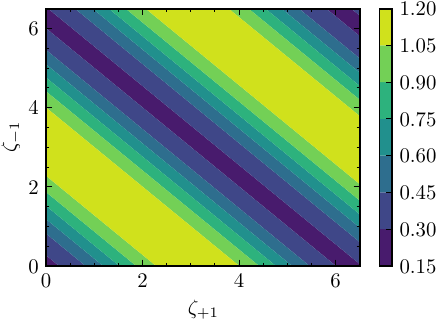}
\end{subfigure}
\hfill
\begin{subfigure}[t]{0.48\textwidth}
\centering
\includegraphics[width=\linewidth]{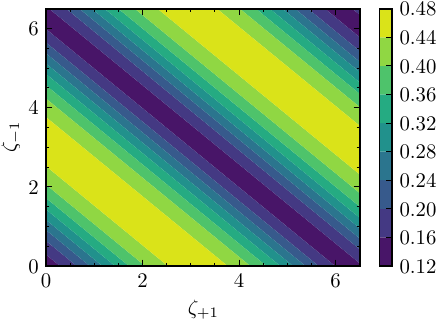}
\end{subfigure}

\caption{Mutual information, $I(A\!:\!B)$ given by Eq.~\eqref{eq:MI}, and negativity, $\mathcal{N}(\Hat{\rho}_{\text{spin}})$ given by Eq.~\eqref{eq:neg}, are shown as functions of the parameter $\zeta_{\pm 1}$ which determines the relative phase between the $M=\pm1$ components in the linear superposition of deuteron spin projections. The value of $\abs{\alpha_2}^2$ is held fixed at 0.0666. Both quantifiers are evaluated for the same spin density matrix to compare the entanglement and correlation behavior under variation of $\zeta_{\pm 1}$.
}
\label{fig:combined_mi_neg_all_m}
\end{figure*}

\begin{figure}[hbt!]
\centering
\includegraphics[width=\linewidth]{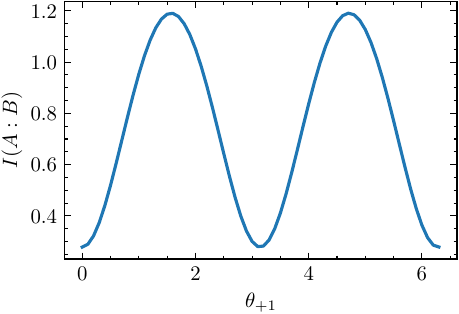}
\caption{Mutual information, $I(A:B)$, given by Eq.~\eqref{eq:MI} is shown as a function of parameter $\theta_{+ 1}$ while fixing the parameter $\abs{\alpha_2}^2 = 0.0666$.}
\label{fig:mu_en_666_0_rotation_single_parm}
\end{figure}

In left panel of Fig.~\ref{fig:combined_mi_neg_all_m}, we present the mutual information as a function of the two-phase parameters while maintaining the free parameter $\abs{\alpha_2}^2 = 0.0666$ (representative of the Woods-Saxon potential). The absence of closed contours suggests that no specific phase value is preferred. Moreover, the straight contour lines indicate that both phases are equally favored. Notably, the maximum correlation occurs when the phases are complementary. This observation is further supported by Table~\ref{tab:table2}, which shows the results of both entanglement quantifiers for specific choices of the phase—namely, $0$ and $\pi$ radians. These phases correspond to the sign $\pm 1$ assigned to the coefficients in Eq.~\eqref{eq:psi_all}, representing the linear superposition of all $M$ projections.
The two specific cases considered in Table~\ref{tab:table2} are the ones where
the signs of the coefficients corresponding to $M=\pm 1$ projections are either different or the same. 
It is evident that for the first case, correlation measures are maximum and remain so for any phase as long as the phases are complementary to each other. Conversely, for the second case, correlation is at a minimum and oscillates between minimum and maximum values if the phases are varied under the constraint that they are equal. Thus, correlation could be measured with the help of single phase parameter, while the other parameter can be considered redundant. One such choice of is given by $\theta_{\pm 1} = (\zeta_{+1} \pm \zeta_{-1})/{\sqrt{2}}$, where $\theta_{\pm1}$ represents a new set of parameters, and $\theta_{-1}$ could be regarded as a redundant parameter. Consequently, the calculated quantifier, i.e., mutual information, is represented by only one parameter, $\theta_{+1}$, as depicted in Fig.~\ref{fig:mu_en_666_0_rotation_single_parm}, which shows the oscillatory behavior of mutual information with phase parameter.

\section{Summary}   \label{sec:conclussion}

We quantified the neutron-proton spin-entanglement in the deuteron bound state.
The bipartite wave function was expressed in the decoupled basis through a repeated utilization of the Clebash-Gordan series, which was followed by a partial tracing to arrive at the reduced density matrix for the spin degrees of freedom.
Since the total angular momentum is $J=1$, we started with the case when its projections could be distinguished in a weak magnetic field.
We found a significant amount of spin-entanglement only in the spin singlet state, ie., for $M=0$.
We further considered a case of the degenerate ground state where the wave function is a linear superposition of all three projections of $J$. The resulting spin density matrix reveal that the nucleons are close to maximally entangled in their spins.

\begin{acknowledgments}
This work is supported by the SERB-DST, Govt.~of India, via project \sloppy{CRG/2022/009359}.
A.K. is supported by NCN SONATA-BIS grant No: 2017/26/E/ST2/01008.
We acknowledge the National Supercomputing Mission (NSM) for providing computing resources of `PARAM Ganga' at IIT Roorkee, which is implemented by C-DAC and supported by MeitY and DST, Govt.~of India.
\end{acknowledgments}

\bibliographystyle{apsrev4-2}
\bibliography{Bibliography}

\cleardoublepage
\onecolumngrid
\appendix

\begin{center}
\large\bf

APPENDICES

$np$ spin correlations in the deuteron ground state

\hrulefill
\end{center}


\section{Spin density matrices} \label{appen:density_matrix_single_proj}

When the system assumes a fixed projection $M$, the reduced density matrix encompassing $np$ spin correlations is given by
\begin{equation}
\Hat{\rho}_{\text{spin}}^{(M)} = \Tr_{R,L,T} \qty(\ket{\Psi^{(M)}}\bra{\Psi^{(M)}}).
\end{equation}
Using Eq.~\eqref{eq:psi_m}, and its corresponding conjugate space representation for $\bra{\Psi^{(M)}}$, we get
\begin{align}
\Hat{\rho}_{\text{spin}}^{(M)} =& \sum_{R_{L}''}\sum_{L'',M_{L''}}\sum_{T'',M_{T''}}\bra{R_{L}''}\bra{L'',M_{L''}}\bra{T'',M_{T''}} \bigl\{\ket{\Psi^{(M)}}\bra{\Psi^{(M)}}\bigr\}\ket{R_{L}''}\ket{L'',M_{L''}}\ket{T'',M_{T''}}
\notag \\[0.5em]
=& \sum_{R_{L}''}\sum_{L'',M_{L''}}\sum_{T'',M_{T''}}\bra{R_{L}''}\bra{L'',M_{L''}}\bra{T'',M_{T''}} \biggl\{\sum_{L}\alpha_{L} \ket{R_{L}}\sum_{M_{L},M_{S}} \mqty( L& S& J  \\  M_{L}& M_{S}& M) \ket{L,M_L} 
\ket{S,M_{S}}\ket{T,M_{T}} 
\notag \\
&  \hspace{1cm} \sum_{L'}\alpha_{L'}^* \bra{R_{L'}}\sum_{M_{L'}',M_{S}'} \mqty( L'& S& J  \\  M_{L'}'& M_{S}'& M) \bra{L',M_{L'}'} \bra{S,M_{S}'}\bra{T,M_{T}} \biggr\}  \ket{R_{L}''}\ket{L'',M_{L''}}\ket{T'',M_{T''}}, \notag \\[0.5em]
=& \sum_{R_{L}''}\sum_{L'',M_{L''}}\sum_{L,L'}\alpha_{L}\alpha_{L'}^* \bra{R_{L}''}\ket{R_{L}} \bra{R_{L'}}\ket{R_{L}''} \sum_{T'',M_{T''}}\bra{T'',M_{T''}}\ket{T,M_{T}} \bra{T,M_{T}}\ket{T'',M_{T''}}  \sum_{M_{L},M_{S}} \mqty( L& S& J  \\  M_{L}& M_{S}& M) 
\notag \\
&  \hspace{1cm} \sum_{M_{L'}',M_{S}'}\mqty( L'& S& J  \\  M_{L'}'& M_{S}'& M) \bra{L'',M_{L''}}\ket{L,M_L} \bra{L',M_{L'}'}\ket{L'',M_{L''}} \ket{S,M_{S}}\bra{S,M_{S}'}  
\notag \\[0.5em]
=& \sum_{R_{L}''}\sum_{L'',M_{L''}}\sum_{L,L'}\alpha_{L}\alpha_{L'}^* \delta_{R_{L}'',R_{L}} \delta_{R_{L'},R_{L}''}  \sum_{T'',M_{T''}} \delta_{T'',T}\delta_{M_{T''}, M_{T}}  \delta_{T,T''}\delta_{M_{T},M_{T''}}  \sum_{M_{L},M_{S}} \mqty( L& S& J  \\  M_{L}& M_{S}& M) \notag \\ 
&  \hspace{1cm} \sum_{M_{L'}',M_{S}'}\mqty( L'& S& J  \\  M_{L'}'& M_{S}'& M) \delta_{L'',L}\delta_{M_{L''},M_L} \delta_{L',L''}\delta_{M_{L'}',M_{L''}} \ket{S,M_{S}}\bra{S,M_{S}'}.
\end{align}
Summing over $R_L'', T''$, and $M_{T''}$, and subsequently over $L''$ and $M_{L''}$, arrive at
\begin{align}
\Hat{\rho}_{\text{spin}}^{(M)} =& \sum_{L'',M_{L''}}\sum_{L,L'}\alpha_{L}\alpha_{L'}^* \delta_{R_{L},R_{L'}} \delta_{T,T}\delta_{M_{T},M_{T}}  \sum_{M_{L},M_{S}}\sum_{M_{L'}',M_{S}'} \mqty( L& S& J  \\  M_{L}& M_{S}& M) \mqty( L'& S& J  \\  M_{L'}'& M_{S}'& M) \notag \\ 
&  \hspace{1cm} \delta_{L'',L}\delta_{M_{L''},M_L} \delta_{L',L''}\delta_{M_{L'}',M_{L''}} \ket{S,M_{S}}\bra{S,M_{S}'}
\notag \\[0.5em]
=& \sum_{L,L'}\alpha_{L}\alpha_{L'}^* \delta_{R_{L},R_{L'}} \sum_{M_{L},M_{S}}\sum_{M_{L'}',M_{S}'} \mqty( L& S& J  \\  M_{L}& M_{S}& M) \mqty( L'& S& J  \\  M_{L'}'& M_{S}'& M) \delta_{L',L}\delta_{M_{L'}',M_{L}} \ket{S,M_{S}}\bra{S,M_{S}'}.  \label{eq:sum_l}
\end{align}
Finally summing over $L'$ and $M_{L'}'$ and using the completeness relation for spin basis, we have
\begin{align}
\Hat{\rho}_{\text{spin}}^{(M)} =& \sum_{L}\abs{\alpha_{L}}^2 \sum_{M_{L},M_{S}}\sum_{M_{S}'} \mqty( L& S& J  \\  M_{L}& M_{S}& M) \mqty( L& S& J  \\  M_{L}& M_{S}'& M) 
\notag \\
&  \hspace{1cm}  \ket{S,M_{S}}  \bra{S,M_{S}'}  \sum_{m_{1},m_{2}}\ket{s_{1},m_{1};s_{2},m_{2}} \bra{s_{1},m_{1};s_{2},m_{2}} \sum_{m_{1}',m_{2}'}\ket{s_{1},m_{1}';s_{2},m_{2}'}\bra{s_{1},m_{1}';s_{2},m_{2}'}  ,
\end{align}
where $\ket{s_{1},m_{1};s_{2},m_{2}}$ represents the uncoupled spin basis. Using the short $\ket{s_{1},m_{1};s_{2},m_{2}} \equiv \ket{m_1,m_2}$, and the fact that $M_L + M_S = M_L + M_S' \implies M_S=M_S'$, we get
\begin{align}
\Hat{\rho}_{\text{spin}}^{(M)} =& \sum_{L}\abs{\alpha_{L}}^2 \sum_{M_{L},M_{S}} \mqty( L& S& J  \\  M_{L}& M_{S}& M)^2 \sum_{m_{1},m_{2}} \mqty( s_1& s_2& S  \\  m_{1}& m_{2}& M_S) \sum_{m_{1}',m_{2}'} \mqty( s_1& s_2& S  \\  m_{1}'& m_{2}'& M_S) \ket{m_{1},m_{2}} \bra{m_{1}',m_{2}'}. \label{eq:rho_m_ab}
\end{align}         

For the system in a generic linear superposition of all $M$ projections we have
\begin{equation}
\Hat{\rho}_{\text{spin}} = \Tr_{R,L,T} \qty(\ket{\Psi}\bra{\Psi}) , \label{eq:psi_all_ab}
\end{equation}
where
\begin{equation}
\ket{\Psi}=\frac{1}{\sqrt{3}}\sum_{M}{e^{i\zeta_M}}\ket{\Psi^{(M)}} , \label{eq:psi_all_ket} 
\end{equation}
which implies  
\begin{align}
\Hat{\rho}_{\text{spin}} =& \sum_{R_{L}''}\sum_{L'',M_{L''}}\sum_{T'',M_{T''}}\bra{R_{L}''}\bra{L'',M_{L''}}\bra{T'',M_{T''}} \biggl\{ \frac{1}{\sqrt{3}}\sum_{M}e^{i\zeta_{M}} \sum_{L}\alpha_{L} \ket{R_{L}}\sum_{M_{L},M_{S}} \mqty( L& S& J  \\  M_{L}& M_{S}& M) \ket{L,M_L}\ket{S,M_{S}} \notag \\
&\ket{T,M_{T}} \frac{1}{\sqrt{3}}\sum_{M'}e^{-i\zeta_{M'}} \sum_{L'}\alpha_{L'}^* \bra{R_{L'}}\sum_{M_{L'}',M_{S}'} \mqty( L'& S& J  \\  M_{L'}'& M_{S}'& M') \bra{L',M_{L'}'} \bra{S,M_{S}'}\bra{T,M_{T}} \biggr\} \ket{R_{L}''}\ket{L'',M_{L''}}\ket{T'',M_{T''}}
\notag\\[0.5em]
=& \sum_{R_{L}''}\sum_{L}\alpha_{L}\bra{R_{L}''}\ket{R_{L}}\sum_{L'}\alpha_{L'}^*\bra{R_{L'}}\ket{R_{L}''} \frac{1}{3}\sum_{M,M'}e^{i\zeta_{M}}e^{-i\zeta_{M'}} \sum_{M_{L},M_{S}} \mqty( L& S& J  \\  M_{L}& M_{S}& M) \sum_{M_{L'}',M_{S}'} \mqty( L'& S& J  \\  M_{L'}'& M_{S}'& M') 
\notag \\ &\sum_{L'',M_{L''}}\bra{L'',M_{L''}}\ket{L,M_L}\bra{L',M_{L'}'}\ket{L'',M_{L''}} \sum_{T'',M_{T''}}\bra{T'',M_{T''}}\ket{T,M_{T}}\bra{T,M_{T}}\ket{T'',M_{T''}} \ket{S,M_{S}}\bra{S,M_{S}'}
\notag \\[0.5em]
=& \frac{1}{3}\sum_{R_{L}''}\sum_{L}\alpha_{L}\delta_{R_{L}'',R_L} \sum_{L'}\alpha_{L'}^*  \delta_{R_{L'},R_{L}''} \sum_{M,M'}e^{i\zeta_{M}}e^{-i\zeta_{M'}} \sum_{M_{L},M_{S}} \mqty( L& S& J  \\  M_{L}& M_{S}& M) \sum_{M_{L'}',M_{S}'} \mqty( L'& S& J  \\  M_{L'}'& M_{S}'& M') 
\notag \\
&\sum_{L'',M_{L''}}\delta_{L'',L}\delta_{M_{L''},M_L}\delta_{L',L''}\delta_{M_{L'}',M_{L''}} \sum_{T'',M_{T''}}\delta_{T'',T}\delta_{M_{T''},M_T}\delta_{T,T''}\delta_{M_T,M_{T''}} \ket{S,M_{S}}\bra{S,M_{S}'}
\notag\\[0.5em]
=& \frac{1}{3}\sum_{M,M'}{e^{i\zeta_{M}}e^{-i\zeta_{M'}}}\sum_{L,L'}{\alpha_{L} \alpha^{*}_{L'}}\sum_{M_L,M_S} \mqty( L & S & J \\ M_{L} & M_{S} & M ) \sum_{M_{L'}',M_S'}\mqty( L' & S & J \\ M_{L'}' & M_{S}' & M')  
\notag \\ 
&  \delta_{L,L'}\delta_{M_{L},M_{L'}'}\delta_{T,T}\delta_{M_T,M_{T}} \ket{S,M_S}\bra{S,M_S'}
\notag\\[0.5em]
=& \frac{1}{3}\sum_{M,M'}{e^{i\zeta_{M}}e^{-i\zeta_{M'}}}\sum_{L}\abs{\alpha_{L}}^2 \sum_{M_L,M_S} \mqty( L & S & J \\ M_{L} & M_{S} & M ) \sum_{M_{L}',M_S'}\mqty( L & S & J \\ M_{L}' & M_{S}' & M') \delta_{M_{L},M_{L'}'}\ket{S,M_S}\bra{S,M_S'}
\notag\\[0.5em]
=& \frac{1}{3}\sum_{M,M'}{e^{i\zeta_{M}}e^{-i\zeta_{M'}}}\sum_{L}\abs{\alpha_{L}}^2 \sum_{M_L,M_S,M_S'} \mqty( L & S & J \\ M_{L} & M_{S} & M ) \mqty( L & S & J \\ M_{L} & M_{S}' & M') \notag \\
& \sum_{m_{1},m_{2}}\ket{s_{1},m_{1};s_{2},m_{2}} \bra{s_{1},m_{1};s_{2},m_{2}} \ket{S,M_S} \bra{S,M_S'}\sum_{m_{1}',m_{2}'}\ket{s_{1},m_{1}';s_{2},m_{2}'} \bra{s_{1},m_{1}';s_{2},m_{2}'} 
\notag \\[0.5em]
=& \frac{1}{3}\sum_{M,M'}{e^{i\zeta_{M}}e^{-i\zeta_{M'}}}\sum_{L}\abs{\alpha_{L}}^2 \sum_{M_L,M_S,M_S'} \mqty( L & S & J \\ M_{L} & M_{S} & M ) \mqty( L & S & J \\ M_{L} & M_{S}' & M') \notag \\
& \sum_{m_{1},m_{2}} \mqty(s_{1} & s_{2} & S \\ m_{1} & m_{2} & M_{S}) \sum_{m_{1}',m_{2}'} \mqty(s_{1} & s_{2} & S 
\notag\\ m_{1}' & m_{2}' & M_{S}') \ket{s_{1},m_{1};s_{2},m_{2}}\bra{s_{1},m_{1}';s_{2},m_{2}'} 
\notag\\[0.5em]
=& \frac{1}{3}\sum_{M,M'}{e^{i\zeta_{M}}e^{-i\zeta_{M'}}}\sum_{L}\abs{\alpha_{L}}^2 \sum_{M_L,M_S,M_S'} \mqty( L & S & J \\ M_{L} & M_{S} & M ) \mqty( L & S & J \\ M_{L} & M_{S}' & M') 
\notag\\
& \sum_{m_{1},m_{2}} \mqty(s_{1} & s_{2} & S \\ m_{1} & m_{2} & M_{S}) \sum_{m_{1}',m_{2}'} \mqty(s_{1} & s_{2} & S \\ m_{1}' & m_{2}' & M_{S}') \ket{m_{1},m_{2}}\bra{m_{1}',m_{2}'}.
\end{align}

\section{Entanglement measures}                \label{appendix:entanglement_measure_formalism}

For pure bipartite system described by a density matrix $\hat \rho_{AB}$, one can quantify the amount of entanglement using the entropy of entanglement, which is defined as von Neumann entropy for any one of the subsystems and is given as~\cite{qi-entropy:PhysRevA.54.3824, qi-entropy:PhysRevLett.78.2275},
\begin{equation}
S(\Hat{\rho}_A) = - \Tr\qty[ \Hat{\rho}_A\log_2(\Hat{\rho}_A) ],    \label{eq:entropy_of_entanglement}
\end{equation}
where $\Hat{\rho}_A = \Tr_B\qty( \Hat{\rho}_{AB} )$ is the reduced density matrix for subsystem $A$.
If $\lambda_i$ are the eigenvalues of $\Hat{\rho}_A$ then above von Neumann’s formula can be equivalently re-expressed as
\begin{equation}
S(\Hat{\rho}_A) = - \sum_{i}\lambda_i \log{\lambda_i}.
\end{equation}

For bipartite mixed states a good correlation quantifier is the mutual information, which is defined as:
\begin{equation}
I(A:B) =  S(\Hat{\rho}_A) + S(\Hat{\rho}_B) - S(\Hat{\rho}_{AB}), \label{eq:MI}
\end{equation}    
where $\Hat{\rho}_A$ and $\Hat{\rho}_B$ are the reduced density matrices for subsystems $A$ and $B$, respsectively.

For separable states, it has been shown that the eigenvalues are positive after partial transposition with respect to one party \cite{qi-neg-ppt:PhysRevLett.77.1413, qi-neg-sep:PhysRevA.58.883, qi-neg-separ:doi:10.1080/09500340008235138, neg-palomi_doi:10.1126/science.1244563}.  This is, in general, referred to as the positive partial transpose (PPT) criterion. If a state is not PPT, one can conclude that the corresponding negative eigenvalues quantify the entanglement in the system. In particular, one defines a computable quantifier of entanglement known as negativity, $\mathcal{N}{\qty(\rho)}$, as a quantitative measure of entanglement that serves as an upper bound for the potential dis-tillable entanglement. It is mathematically given by:
\cite{qi-neg-gen:PhysRevA.65.032314}.
\begin{equation}
\mathcal{N}{\qty(\Hat{\rho}_{AB})} =  \frac{\norm*{\Hat{\rho}^{\Gamma_A}_{AB}}_1 - 1}{2},
\label{eq:neg}
\end{equation}
where $\norm*{\Hat{\rho}^{\Gamma_A}_{AB}}_1$ is the trace norm of the state after partial transposition with respect to subsystem $A$. Variations of this measure, such as log-negativity, are widely used across quantum information science.

\end{document}